\documentclass[conference, table]{IEEEtran}
\IEEEoverridecommandlockouts
\AtBeginDocument{%
  }

\usepackage{algorithm}
\usepackage[noend]{algpseudocode}
\usepackage{pgf}
\usepackage{float}  
\usepackage{dsfont}
\usepackage{subcaption}
\usepackage{multirow}
\usepackage{tabularx}
\usepackage{pgfmath}
\usepackage{lipsum}
\usepackage{comment}
\usepackage{url}
\usepackage{hyperref}
\usepackage{booktabs}
\usepackage{amsmath}

\usepackage{makecell}

\makeatletter
\newcommand{\linebreakand}{%
  \end{@IEEEauthorhalign}
  \hfill\mbox{}\par
  \mbox{}\hfill\begin{@IEEEauthorhalign}
}
\makeatother

\begin{document}

\title{BroadGen: A Framework for Generating Effective and Efficient Advertiser Broad Match Keyphrase Recommendations}

\author{\IEEEauthorblockN{Ashirbad Mishra}
\IEEEauthorblockA{\textit{eBay Advertising} \\
\textit{eBay Inc.}\\
San Jose, California, USA \\
ashirmishra@ebay.com}
\and
\IEEEauthorblockN{Jinyu Zhao}
\IEEEauthorblockA{\textit{eBay Advertising} \\
\textit{eBay Inc.}\\
San Jose, California, USA \\
jinyzhao@ebay.com}
\and
\IEEEauthorblockN{Soumik Dey}
\IEEEauthorblockA{\textit{eBay Advertising} \\
\textit{eBay Inc.}\\
San Jose, California, USA \\
sodey@ebay.com} \\
\linebreakand
\IEEEauthorblockN{Hansi Wu}
\IEEEauthorblockA{\textit{eBay Advertising} \\
\textit{eBay Inc.}\\
San Jose, California, USA \\
hanswu@ebay.com}
\and
\IEEEauthorblockN{Binbin Li}
\IEEEauthorblockA{\textit{eBay Advertising} \\
\textit{eBay Inc.}\\
San Jose, California, USA \\
binbli@ebay.com}
\and
\IEEEauthorblockN{Kamesh Madduri}
\IEEEauthorblockA{\textit{Dept. of Computer Science} \\
\textit{Pennsylvania State University}\\
University Park, Pennsylvania, USA \\
madduri@psu.edu}
}
\maketitle

\begin{abstract}
In the domain of sponsored search advertising, the focus of \textit{Keyphrase recommendation} has largely been on exact match types, which pose issues such as high management expenses, limited targeting scope, and evolving search query patterns. Alternatives like \textit{Broad} match types can alleviate certain drawbacks of exact matches but present challenges like poor targeting accuracy and minimal supervisory signals owing to limited advertiser usage. This research defines the criteria for an ideal broad match, emphasizing on both efficiency and effectiveness, ensuring that a significant portion of matched queries are relevant. We propose \textit{BroadGen}, an innovative framework that recommends efficient and effective broad match keyphrases by utilizing historical search query data. Additionally, we demonstrate that BroadGen, through token correspondence modeling, maintains better query stability over time. BroadGen's capabilities allow it to serve daily, millions of sellers at eBay with over 2.5 billion items.
\end{abstract}


\begin{IEEEkeywords}
Information Retrieval, Sponsored search advertising, Keyword Recommendation
\end{IEEEkeywords}


\maketitle

\section{Introduction}
\label{s:intro}
\begin{figure}[t]
\centering
\begin{subfigure}{8.25cm}
\centering
\includegraphics[width=\linewidth]{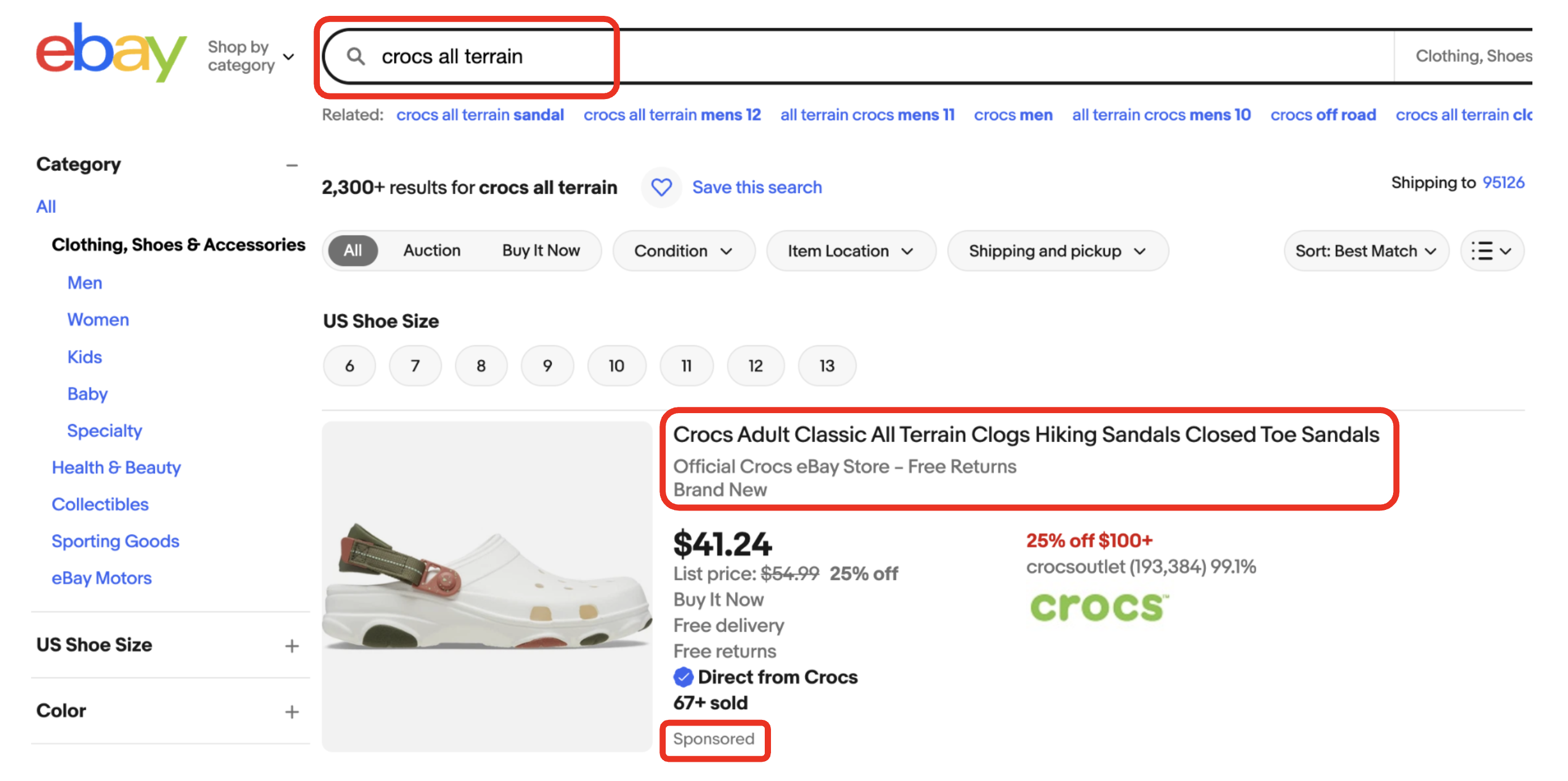}
  \caption{View of Buyer side.}
  \label{fig:1a}
\end{subfigure}
\begin{subfigure}{8.25cm}
\centering

\includegraphics[width=\linewidth]{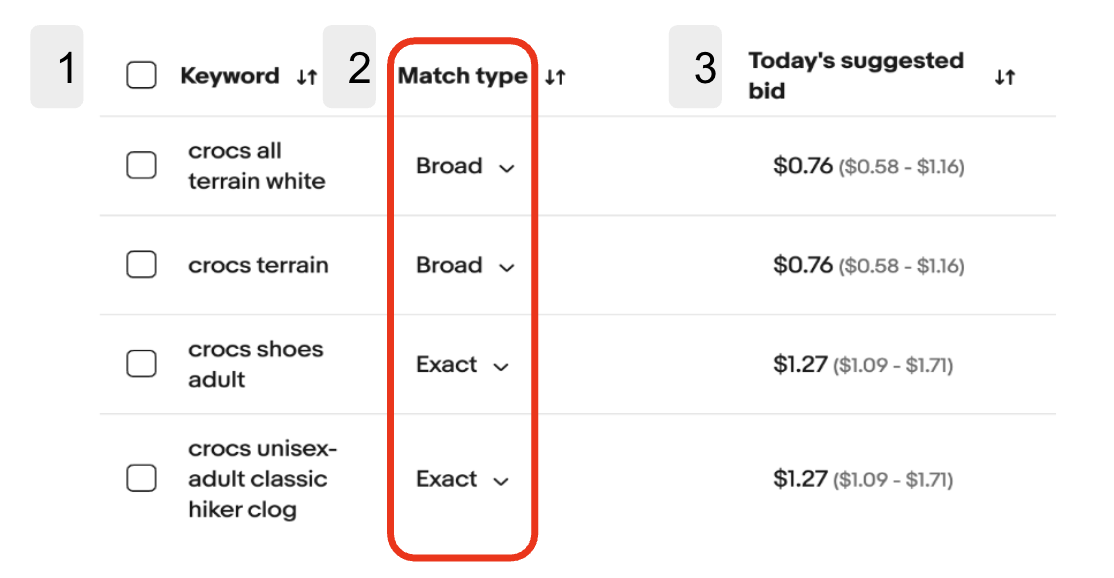}
\caption{View of Seller side.}
\label{fig:1b}
\end{subfigure}
\caption{A view of buyer (a) and seller side (b) of the keyphrase recommendation system for eBay advertising (manual targeting).}
\vspace{-6mm}
\label{fig:screenshot}
\end{figure}

Advertisers or sellers on e-commerce platforms are recommended keyphrases by the platform to increase their item visibility depending on buyer searches as shown in Figure~\ref{fig:1a}. The sellers bid on relevant recommended keyphrases as well as their own custom keyphrases to increase their sales (Figure~\ref{fig:1b}). \textit{Keyphrase}  recommendation is the subject of a lot of research and well studied in several works~\cite{unidec_www25,mishra2025graphexgraphbasedextractionmethod,ashirbad-etal-2024,dahiya2021deepxml,triangular_bidgen,renee_2023,Dahiya23b}, due to their challenging research and industrial importance. We term the recommendation to the seller as \textit{Keyphrase} and the buyer search text as \textit{Query}. Additionally, the platform \textit{matches} the (recommended) keyphrases bid by the sellers to the buyer search queries~\cite{dey2025middlemanbiasadvertisingaligning}.

Each keyphrase is also tagged with a \textit{Match Type}, which is used during the matching process; for this work we define them as:
\begin{itemize}
    \item \textit{Exact} match type - The keyphrase is matched exactly to the query accommodating different cases, plurals, contractions, and similar substitutions handled by tokenization.
     \item \textit{Phrase} match type - It is more flexible by allowing modification to the keyphrase by in-position token expansion but maintaining the token order.\label{def:phrase}
    \item \textit{Broad} match type - In addition to the criteria of phrase match it is more flexible by disregarding order constraints.\label{def:broad}
\end{itemize}

\begin{figure}[t]
    \centering
    \includegraphics[width=\linewidth]{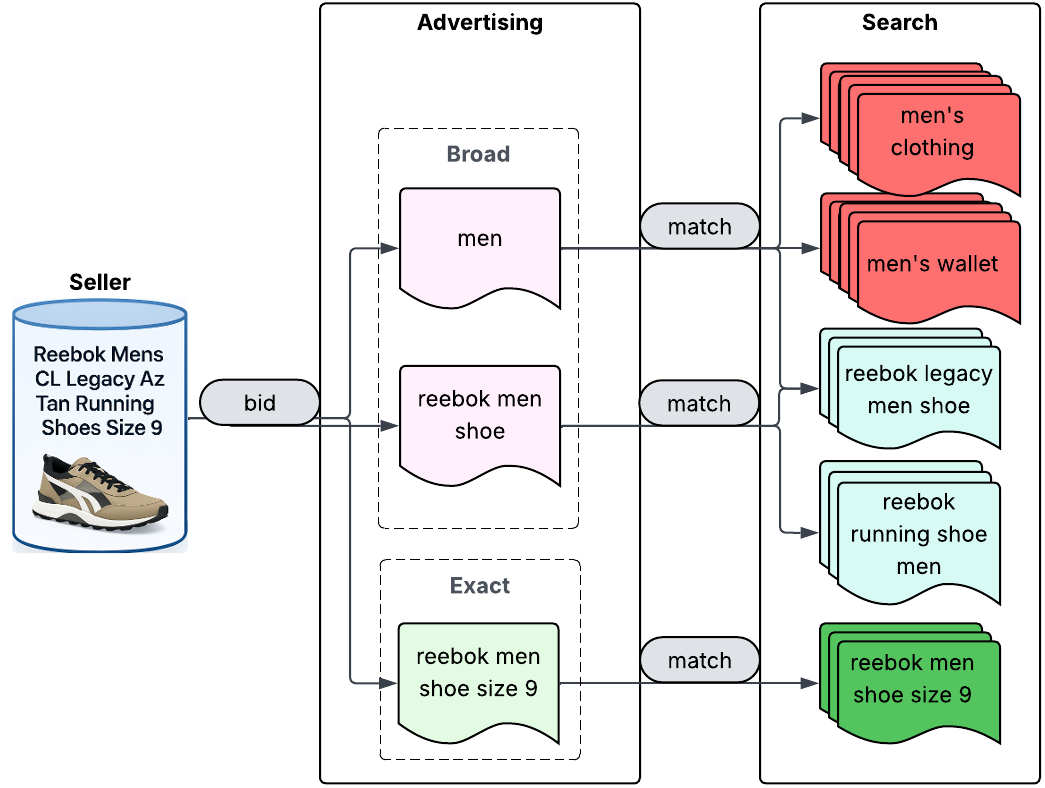}
    \caption{Architecture for matching seller's keyphrases with different match types to buyer queries.}
    \label{fig:matching_architecture}
    \vspace{-3mm}
\end{figure}

For example, if a query is \texttt{``reebok men's shoes size 9''}, then the exact matching keyphrases can be \texttt{``reebok men shoes size 9''} or \texttt{``Reebok men's shoe size 9''}. The phrase match can be \texttt{``reebok men shoe''} while broad match can be \texttt{``reebok shoe men''}. More examples of how different match types keyphrases for an item are matched to queries are shown in Figure~\ref{fig:matching_architecture}. The platform can recommend a keyphrase with different match types to the sellers. 

Previous research has focused on recommending exact match type keyphrases, popularly framing it as an Extreme Multi-Label Classification (XMC) \cite{Bhatia16, agrawal2013multi}, where given an input item, they determine relevance to the labels which in this case are the buyer queries or keyphrases. Match types such as phrase and broad match keyphrases can alleviate problems associated with exact match type recommendations (see Section~\ref{ss:import_match_types}). However, their recommendation strategies are under-researched as they are unlike exact match recommendations. Therefore, in this paper, we focus on other match type keyphrases by formulating the problem, designing solutions and providing insights learned in production.


\subsection{Problems with Exact Match Type}
\label{ss:import_match_types}
\subsubsection{Shifting label distributions}
\label{sss:shift_label_dist}
Analysis reveals that in sponsored search, queries adhere to the 80/20 rule where 20\% of queries generate 80\% of search volume leaving 80\% as long-tail queries~\cite{ZHANG2014508}, which is prominent even in e-commerce platforms like eBay. These long-tail queries are low in volume, transient, and rarely reappearing monthly. This transient nature limits the effectiveness of traditional exact match algorithms~\cite{unidec_www25,mishra2025graphexgraphbasedextractionmethod,ashirbad-etal-2024,dahiya2021deepxml, dey2025judgejudgeusingllm}, as most labels won't have future search volume. Due to the shifting dynamics of long-tail queries, predicting them individually as exact match keywords is nearly infeasible at time of inference. Instead, using other match types addresses this issue without explicit demand forecasting. Broad match keyphrases inherently address this challenge by capturing a wider set of related queries, thus effectively mitigating the unpredictability and transient nature of long-tail search patterns.

\subsubsection{Management costs and Sales}
\label{sss:mang_costs_revenue}
 Managing sponsored search campaigns demands substantial effort from sellers, compounded by the extensive set of keyphrases requiring bids~\cite{ZHANG2014508,Amaldoss-2016,YANG2021113491}. Sellers prefer concise recommendations due to their perception and business demands~\cite{mishra2025graphexgraphbasedextractionmethod, dey2025judgejudgeusingllm}. Other match type keyphrases are beneficial for targeting a wide array of exact queries, thereby reducing seller costs and involvement. Research suggests that when curation accuracy is high and seller bids remain competitive, broad match keyphrases yield higher sales~\cite{YANG2021113491}.


\subsection{Challenges for Other Match Types}
\label{ss:challenges}
Though non-exact match types are lucrative there are several challenges associated with recommending them.

\subsubsection{Ground Truth Hurdles}
\label{sss:lack_ground_truths}
Exact match keyphrases can be directly recommended by modeling on click data logs, since they correspond precisely to buyer queries. In contrast, phrase and broad match keyphrases are more difficult to model due to the sparsity and bias of available training signals. It suffers from sparsity as the performance information of phrase/broad match types are only collected when the seller manually chooses them. Although the sellers' previous picks of successful broad match keyphrases could seem like a good basis for ground truth, they can be heavily affected by the seller set bid amounts and auctioning dynamics~\cite{bid_auctioning07}, thus making seller-based preference data a bad choice for modeling keyphrase recommendations \cite{dey2025judgejudgeusingllm}. 

Several biases with click data have been well studied~\cite{mishra2025graphexgraphbasedextractionmethod,surveybias,sampleselectionbias,rec4ad,dey2025middlemanbiasadvertisingaligning}, these impact any form of supervised training for recommendation. The limited availability of ground truth data leaves a significant portion of the feature space unmarked, requiring the incorporation of counterfactual data so models can learn significant patterns. Moreover, missing label biases~\cite{MNAR} affect substantial counterfactual negative signals, as the absence of impressions or clicks doesn't mean that the match type keyphrases are irrelevant. 

Consequently, the validity and dependability of historical performance metrics for assessing these recommendations become dubious. So, unlike exact match recommendations, which can be easily evaluated using classification metrics such as precision and recall score in a traditional setting (using ground truths), evaluating other match types presents significant challenges.

\subsubsection{Active Targeting}
\label{sss:active_targeting}
Keyphrases of non-exact match type recommended to sellers should be reachable to queries that are searched by buyers. Generated broad/phrase match keyphrases which are open vocabulary cannot reach such queries, though relevant to the item are considered futile leading to wastage of seller bids.\footnote{See discussion of Out-of-Vocabulary models~\cite{PUSL, simig-etal-2022-open, zest-xml,triangular_bidgen} and targeting in Graphex~\cite{mishra2025graphexgraphbasedextractionmethod}.} This is exacerbated by the transient queries mentioned in Section~\ref{sss:shift_label_dist} that require keyphrases to target queries that buyers search for over time. Thus, there should be a way of evaluating the performance of broad matches on transient queries.

\subsubsection{Complexity of Reach}
\label{sss:reach_match_types}
If a phrase/broad match keyphrase can be extended to a query using the definition in Section~\ref{s:intro}, then we term it as \textit{Reach}. Theoretically, a keyphrase can reach a larger set of queries, assuming there is a finite number of actively buyer search queries. The broader the keyphrase the larger the reach. For example as shown in Figure~\ref{fig:matching_architecture}, a generic broad type keyphrase \texttt{``men''} can reach thousands of queries pertaining  to shoes, clothing, or accessories. Whereas a more specific \texttt{``reebok men shoe''} can reach only queries related to the specific brand, gender and accessory.

Although the broader keyphrase is beneficial to the seller by reaching a larger query set, it can lead to inefficient curation by flooding the matching platform illustrated in Figure~\ref{fig:matching_architecture}. The matching for the more generic \texttt{``men''} keyphrase will include a larger number of queries which are mostly irrelevant and increase computational costs for the platform.
Also, due to the matching complexity of larger reach for the generic keyphrase, there is a high possibility of mismatching the seller's item to an unprofitable buyer search query. This suboptimal targeting not only reduces the platform’s revenue, but also diminishes the seller’s return on ad spend (ROAS). Ideally, we want the broad match keyphrase to be closely matched with the reachable queries so that it takes less modifications to match the queries exactly.

\subsubsection{Execution Requirements}
\label{sss:execution_requirements}
Executing billions of item and query inferences within a few hours is paramount for the timely completion of daily tasks. Given the high memory demands, single-node data aggregation for inference poses a failure risk, while multi-node management elevates engineering expenses. Models need frequent updates for continuous integration of new data points, as discussed in Section~\ref{sss:active_targeting}. Deep learning networks like Large Language Models (LLMs) face scrutiny for high execution costs, notably due to their training and inference latency~\cite{kaddour2023,mishra2025graphexgraphbasedextractionmethod,10272546,MLSYS2023_LLMscaling}. Leading LLMs often rely on GPUs, raising acquisition costs and affecting both advertisers' and platforms' profit margins.

\subsection{Scope and Contributions}
\label{ss:scope_contri}
In this work we only focus on \textit{Broad} matches instead of \textit{Phrase} matches~\cite{ebayMatchTypes,googleMatchTypes} due to their wide applicability. We use our definition in Section~\ref{s:intro} to streamline the evaluation; the actual implementation of the matching is platform-specific. We aim to tackle the challenges mentioned in Section~\ref{ss:challenges}, which we briefly describe here. We present a novel framework \textit{BroadGen} for recommending broad match keyphrases to sellers and evaluate its performance. Due to the lack of ground truths per item (see Challenge~\ref{sss:lack_ground_truths}), we incorporate relevant buyer queries as input (see Section~\ref{ss:setup_datasets}) to determine the effective keyphrases for an item. This also helps us target active queries for Challenge~\ref{sss:active_targeting}. The Challenge~\ref{sss:reach_match_types} implies that a good broad match keyphrase should be neither too generic nor too specific, so that it can achieve a balance between reach and effectiveness. A \textit{``too specific''} broad match keyphrase will behave like an exact match keyphrase which defeats the purpose. Our approach emphasizes avoiding overly broad suggestions, particularly in Section~\ref{ss:impl_disc}, with the aim of mitigating inefficiencies within both our systems and the business metrics of sellers. Our framework has been designed to be light and fast with minimal infrastructure costs and is easily deployable in a distributed manner (Challenge~\ref{sss:execution_requirements}) to cater to millions of sellers with an inventory of over 2.5 billion items daily.
We also develop a unique evaluation methodology described in Sections~\ref{ss:setup_datasets} and ~\ref{ss:metrics} that enables us to calculate the balance in effectiveness and efficiency of the recommended broad match keyphrases in relation to the search queries it will be matched to. It allows for absence of ground truths and assesses performance over time, emphasizing their adaptability and temporal resilience (Challenge~\ref{sss:active_targeting}).

\section{Related Work}
\label{s:related_work}

Although broad match keyphrases are a crucial component, the research dedicated to specifically recommending novel broad match keyphrases is sparse. Traditionally, sellers have been responsible for selecting match types, with prior studies~\cite{SCHOLZ201996} concentrating on creating keyphrases suitable for exact matches. A simplistic recommendation strategy has involved a rule-based \textit{Match Type Curation}, which applies heuristics to assess whether an exact match type keyphrase, obtained through the current recall mechanism, has historically succeeded as a broad match keyphrase. This approach, however, faces intrinsic challenges: without historical data indicating that a seller has previously chosen the keyphrase as a broad match type --- this method is unable to suggest the keyphrase due to a lack of pertinent performance indicators. Note that in this method the keyphrase is a query in itself.

\begin{figure}[t]
    \centering
    \includegraphics[width=\linewidth]{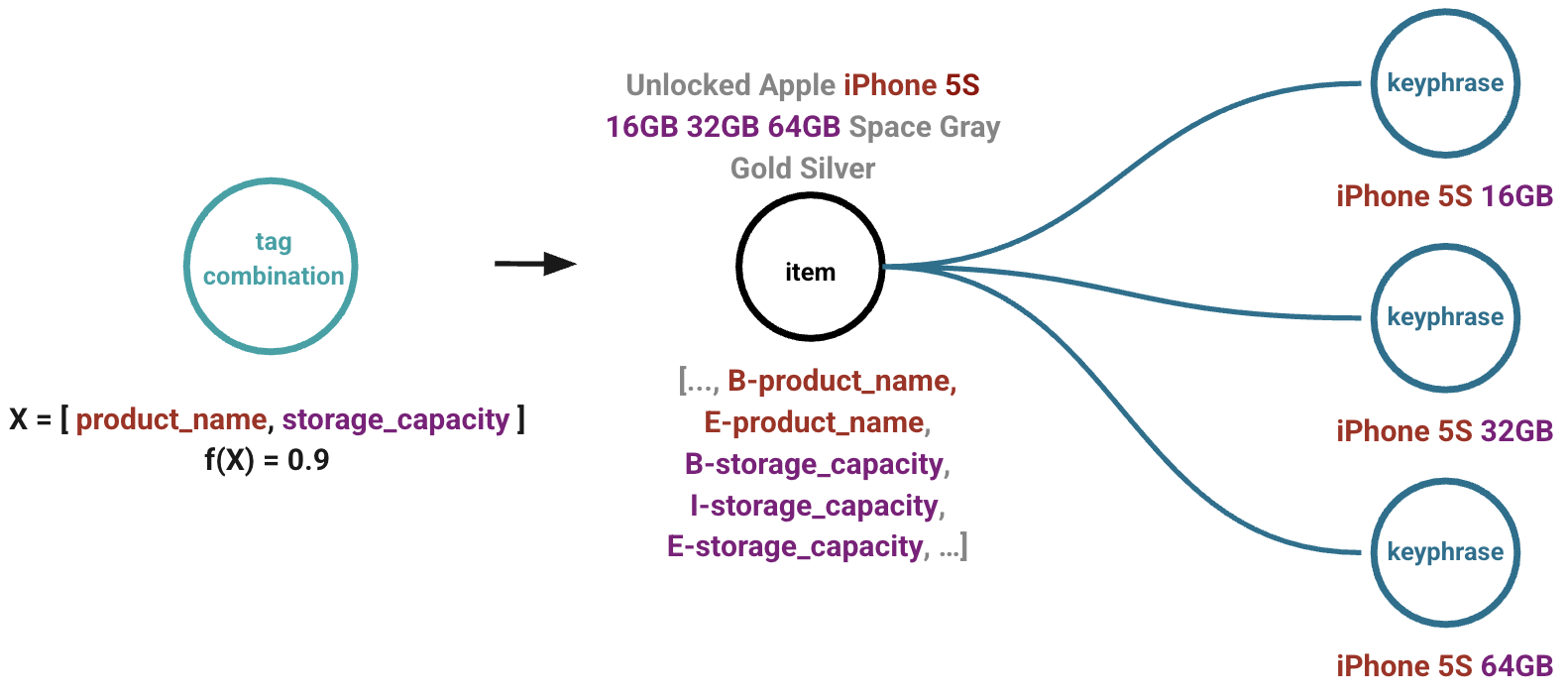}
    \caption{Example recommendation of keyphrases by the Transfusion using NER tags.}
    \label{fig:transfusion_diagram}
    \vspace{-3mm}
\end{figure}

Some studies have focused on optimizing the matching of keyphrases to queries rather than effecting the sellers/advertiser's selection of keyphrases. The BB-KSM model~\cite{LI2022101209} is a stochastic model that utilizes Markov Chain Monte Carlo (MCMC) to initially estimate keyphrase metrics such as impressions and CTR across match types, filling in incomplete data. Subsequently, it designates a match type to each keyphrase based on these metrics with the aim of maximizing a profit objective under budgetary constraints, employing a branch-and-bound algorithm for efficient decision space exploration. Similar optimization strategies have been documented for match types~\cite{dar-etal-2009}, which recommend using buyer search queries as broad match keyphrases rather than introducing new broad keyphrases. However, with the increasing variety of items and queries in e-commerce, this approach may be inefficient and expensive. The introduction of new queries and changing distributions (refer to Section~\ref{sss:shift_label_dist}) further reduces profitability. Previous studies~\cite{radlinski_broder_SIGIR08} apply query substitution methods~\cite{rosiejones_querydeletion_sigir03,jones_rey_madani_www06} to connect broad keyphrases with buyer queries by preprocessing frequently searched queries and utilizing predefined substitution mappings.

\textit{Transfusion} is an eBay's match type generative model developed based on its in-house NER (Named Entity Recognition) model \cite{xin-etal-2018-learning, palen-michel-etal-2024-queryner} to recommend broad match type keyphrases especially for cold start items. The model generates keyphrases for a specified item utilizing the tokens present in its title. It ensures that these keyphrases align with popular and historically high-performing NER tag combinations within a particular meta category. For instance, as shown in Figure~\ref{fig:transfusion_diagram}, the model identifies that the NER tag combination of \textit{product\_name} and \textit{storage\_capacity} is popular in this category based on the scoring function $f(X)$. The algorithm then construct 3 keyphrases: \texttt{``iPhone 5S 16GB''}, \texttt{``iPhone 5S 32GB''}, \texttt{``iPhone 5S 64GB''} based on the tokens corresponding to this NER tag combination from the item title. In this work, for comparison, we use a proxy for the rule-based match type curation and the Transfusion model against our model \textit{BroadGen}.

\section{Methodology}
\label{s:methodology}

\subsection{Terms, Notations and Formulation}
\label{ss:terms_notations_formulation}

Our framework requires a set of existing (buyer search) queries denoted as $Q$ and the item title $t$ for each item to recommend a set of keyphrases denoted as $K$. The goal is for the keyphrase set $K$ to be broad matched to the queries $Q$. Each query $q \in Q$, each keyphrase $k\in K$ and the title $t$ are in the form of a list of tokens.\footnote{A proprietary tokenization scheme handles normalization and stop word removal.} We formulate the problem as \textit{String Clustering} on the set of query strings ($Q$), to exploit the commonality within a cluster of queries and generate a representation for each cluster. The underlying intuition is that the cluster representation serves as the broad match keyphrase capable of addressing both the queries within the set and prospective queries beyond it. Our framework has three cores, \textit{Input generation} (\ref{ss:input_gen}) which generates the input matrix from the set of input queries. The \textit{Clustering core} (\ref{ss:clustering_core}) which groups the queries using the input matrix. The \textit{Representation generation} (\ref{ss:cluster_repr_order}) constructs a representative ordered string for each cluster.

\subsection{Input Generation}
\label{ss:input_gen}
Two input matrices are constructed, Anchor Similarity Matrix $A^{|Q|\times |t|}$ and Query-Query Similarity Matrix $S^{|Q|\times |Q|}$, where $|\cdot|$ determines cardinality. In constructing $A$, a $|t|$-dimensional vector is generated for each query. This vector comprises of binary values, where 1s and 0s indicate the query token's presence in the title. The construction of $S$ is done by computing the common token count between each pair of queries in $Q$. Each row of $S$ can be considered as a $|Q|$-dimensional vector that represents each query depending on similarity with other queries. Each vector of matrix $S$ is normalized by dividing by its maximum, while each vector of the matrix $A$ is normalized by its sum. The Query-Query Similarity matrix $S$ is transformed into a distance matrix by $D_{ij}=1-S_{ij}$. The core combines both the matrices $A$ and $S$ as shown in

\begin{algorithm}[t]
\begin{algorithmic}[1]
\Require Input matrix $\tilde{Z}$, threshold $\tilde{r}$ and required number of clusters $|K|$
\Ensure List of clusters
\Function{Retrieve}{$R,\tilde{Z},\tilde{r},|K|$}
\State $C\gets R(\tilde{Z},\tilde{r})$ 
\For {$|C|< |K|$}
    \State $\tilde{r}\gets\tilde{r}-\epsilon$ \label{alg:ratio}
    \State $C\gets R(\tilde{Z},\tilde{r})$
\EndFor
\If{$|C|>|K|$}
    \State $C\gets pick(C,|K|)$ \Comment{Pick largest clusters.}\label{alg:sort}
\EndIf
\State \textbf{return} $C$
\EndFunction
\end{algorithmic}
\caption{Retrieve clusters generated by $R(\cdot,\cdot)$}
\label{alg:clustering_core}
\end{algorithm}

\begin{equation}
\label{eq:matrix}
    Z=\left[A\;|\;D\right],\quad\hat{Z}=\frac{Z}{||Z||_2}
\end{equation}

 The normalized augmented matrix $\hat{Z}$, provides a vector representation for each query anchoring in relation to the title and acts as a query distance vector when there is insufficient information in the anchor representation. 

\subsection{Clustering Core}
\label{ss:clustering_core}
The core functionality starts with computing the pairwise distance between each pair of vectors $z_i,z_j\in \hat{Z}$ as demonstrated in 
\begin{equation}
\label{eq:ward}
    \tilde{Z}=1-\frac{z_i\cdot z_j}{||z_i||_2\cdot||z_j||_2}
\end{equation}


Now, with the pairwise distance matrix $\tilde{Z}$, the core performs agglomerative clustering~\cite{müllner2011modernhierarchicalagglomerativeclustering} by constructing linkage distances with ward variance minimization~\cite{Ward01031963} and optimal leaf ordering~\cite{optimalLeaforder2001}. We denote the clustering function as $R(\tilde{Z},\tilde{r})$ which outputs a set of clusters $C={c_1,c_2,...}$, where each cluster $c_i$ groups similar queries together. While $\tilde{r}$ is a crucial hyper-parameter called the threshold ratio that determines how many clusters are formed.\footnote{We use the \textit{inconsistent} criterion that compares the linkage distance of current and previous levels of the hierarchy.} A fixed value of $\tilde{r}$ can lead to more or less number of clusters $|C|$ than the requirement $|K|$. The Algorithm~\ref{alg:clustering_core} shows how the required number of clusters is retrieved, through decreasing the threshold by epsilon (line~\ref{alg:ratio}) when there are less clusters. If there are more clusters than required, it picks the top $|K|$ clusters (line~\ref{alg:sort}) in decreasing order of the cluster size.

\subsection{Representation Generation}
\label{ss:cluster_repr_order}
A representation for each cluster in $C$ depicts a keyphrase that can be extended to queries in the cluster. In the final core, the representation is determined as the set of tokens common among the tokenized queries within each cluster.  At times, clusters can be large leading to single or no tokens (in absence of commonality). In these situations, an optimal subset of the cluster is identified whose representation is more than one token. The common tokens from each cluster need to be ordered before concatenating into a string. We employ a graph-based approach to infer an ordering of the common tokens based on their relative positions across the queries. A graph is constructed with all tokens as the vertices; for each co-occurring pair of tokens a directed edge is added indicating local precedence. A topological sort~\cite{Tarjan_1976} of the graph using a depth-first search algorithm produces a consistent global ordering of the tokens. This method of determining precedence is rooted in prior work, such as sentence ordering task~\cite{prabhumoye-etal-2020-topological}. For each cluster, the common tokens are arranged following the global order and represented as a string.

\begin{figure}[t]
    \centering
    \includegraphics[width=\linewidth]{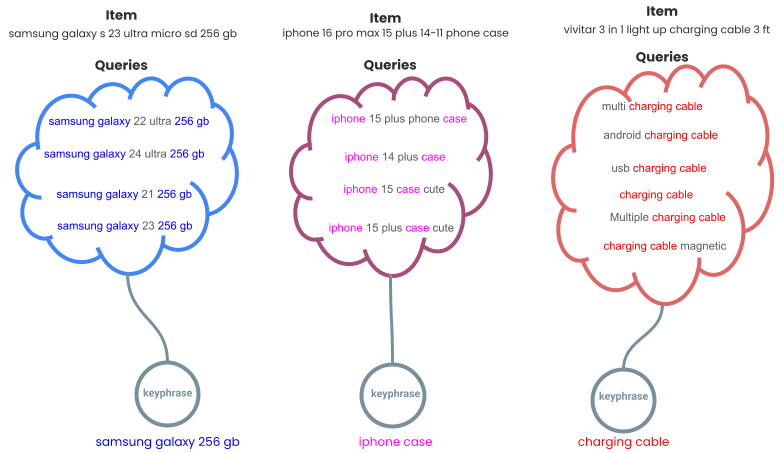}
    \caption{Example construction of keyphrases by BroadGen from clusters of queries. Illustration shows one cluster per item.}
    \label{fig:our_model_diagram}
    \vspace{-2mm}
\end{figure}

\subsection{Implementation Details}
\label{ss:impl_disc}

We show an example of our model on three different items in Figure~\ref{fig:our_model_diagram}. For each item, only one cluster of queries is shown along with the keyphrases that are generated. \footnote{This is just for illustration, there can be multiple clusters per item depending on the number of input queries.} Aligning with our goals described in Section~\ref{sss:execution_requirements}, we develop the framework in Python to enable its integration into eBay's in-house Spark system. This off-loads the overhead of data transfer and parallelism to the spark system, making it feasible to scale for billions of items while mitigating engineering costs. The cores of the framework are designed to be fast and efficient. The asymptotic time complexity of all the cores is $O(|Q|^2)$ which might seem large but $Q$ is per item which is typically in the thousands.

The framework's recommendations are adaptive to the input queries --- depending on number and quality of queries, the recommendations change significantly. Clustering the queries helps us optimize the reach of a particular keyphrase by limiting it's tokens to emerge from the cluster which assists to some extent in identifying \textit{relevant reach} (to be discussed later in Section \ref{metric:relevant_coverage}). To obtain the cluster representation, we impose specific constraints to prevent the inclusion of generic broad keyphrases from the cluster. If the retrieved representation consists of a single token, we instead use a subset of the cluster for retrieval. The subset is chosen from all possible combinations of size $|C|-1$ where the retrieved representation is more than a single token.

\subsection{Explainability}

Since BroadGen is based on token correspondence between candidate exact match keyphrases (queries) with an underlying pre-query population coming from historical buyer interaction, it is explainable. Given a BroadGen keyphrase we would be able to provide the exact clustering of the input pre-queries that were considered and the token distribution within the cluster, summarizing the differences between the queries as illustrated in Figure~\ref{fig:our_model_diagram}.


\section{Experimentation}
\label{s:evaluation}

\subsection{Setup}
\label{ss:setup_datasets}

\subsubsection{OOT Formulation}

To evaluate the effectiveness of broad match keyphrases in addressing the challenge of shifting label distribution, we assess their generalized performance in an \textit{Out-of-Time (OOT)}~\cite{PUSL} setting \cite{OOD2,yu2024surveyevaluationoutofdistributiongeneralization, liu2024robustnessgenerativeinformationretrieval}. Specifically, the algorithms for generating broad match keyphrases use as input the historical queries searched by buyers prior to time $T$, termed as \textit{Pre-queries} and evaluated against the query distribution observed in the subsequent $T+30$ day period called as \textit{Post-queries}. 

\subsubsection{Data Augmentation}
The primary limitation of BroadGen's methodology lies in its dependency on historical interaction data between items and queries for the generation of keyphrases. Although the algorithm functions effectively under warm start scenarios, it encounters difficulties with new or \textit{``cold''} items that have limited interaction data available. To address this shortcoming, our approach incorporates input from state-of-the-art models deployed at eBay to provide precise keyphrase recommendations for cold-start items~\cite{ashirbad-etal-2024, mishra2025graphexgraphbasedextractionmethod}, thereby bolstering the performance of our algorithm when dealing with cold-start conditions.

\begin{table}[t]
\centering
\begin{tabular}{l|ccc}
\toprule
\textbf{Category} & \textbf{\# Items} & \textbf{\# Unique Queries} & \textbf{\#Queries/\#Items} \\ \midrule
CAT\_1            &  210M             &   17.2M             &  15.12                    \\
CAT\_2            &  13.8M            &   12.2M             &  40.60                     \\
CAT\_3            &  14.9M            &   7.5M              &  21.52                     \\
CAT\_4            &  4.3M             &   2.8M              &  29.80                      \\
\bottomrule
\end{tabular}
\caption{Number of items, queries and average queries per item in product categories from eBay.}
\label{tab:cat_details}
\vspace{-2mm}
\end{table}

\subsubsection{Datasets} We present results on 4 meta-categories of eBay, CAT\_1 to CAT\_4, each representing a group whose size is based on number of items and queries. The category sizes are largest from CAT\_1 to CAT\_4, the detailed statistics for each category is presented in Table \ref{tab:cat_details}. For each category, a buyer search query is associated with an item if the \textit{buyer clicked} on it (see curated datasets in Graphite~\cite{ashirbad-etal-2024}). Each item's future query distribution $Q$ is limited to 1000 aggregated queries. Each of these 1000 queries has at least 1 token overlap with the item's title. 

\definecolor{goodgreen}{RGB}{0, 90, 0}
\addtolength{\tabcolsep}{-1.5pt}
\begin{table*}[t]
\centering
\small
\begin{tabular}{
>{\columncolor[HTML]{FFFFFF}}l ||cc
>{\columncolor[HTML]{FFFFFF}}c c||cccc||cccc}
\hline
\cellcolor[HTML]{FFFFFF}                                  & \multicolumn{4}{c||}{\cellcolor[HTML]{C0C0C0}\textbf{Precision}}                                                                                              & \multicolumn{4}{c||}{\cellcolor[HTML]{C0C0C0}\textbf{Recall}}                                                                                                          & \multicolumn{4}{c}{\cellcolor[HTML]{C0C0C0}\textbf{F1}}                                       \\ 
\multirow{-2}{*}{\cellcolor[HTML]{FFFFFF}\textbf{Models}} & \cellcolor[HTML]{FFFFFF}\textbf{CAT\_1} & \cellcolor[HTML]{FFFFFF}\textbf{CAT\_2} & \textbf{CAT\_3}                & \cellcolor[HTML]{FFFFFF}\textbf{CAT\_4} & \cellcolor[HTML]{FFFFFF}\textbf{CAT\_1} & \cellcolor[HTML]{FFFFFF}\textbf{CAT\_2} & \cellcolor[HTML]{FFFFFF}\textbf{CAT\_3} & \cellcolor[HTML]{FFFFFF}\textbf{CAT\_4} & \cellcolor[HTML]{FFFFFF}\textbf{CAT\_1} & \textbf{CAT\_2} & \textbf{CAT\_3} & \textbf{CAT\_4} \\
\midrule
\textbf{PreQuery Model} &
\cellcolor{goodgreen!7!white}\textcolor{black}{0.48} & \cellcolor{goodgreen!15!white}\textcolor{black}{0.52} & \cellcolor{goodgreen!0!white}\textcolor{black}{0.50} & \cellcolor{goodgreen!5!white}\textcolor{black}{0.63} &
\cellcolor{goodgreen!76!white}\textcolor{white}{0.32} & \cellcolor{goodgreen!51!white}\textcolor{black}{0.38} & \cellcolor{goodgreen!52!white}\textcolor{black}{0.32} & \cellcolor{goodgreen!32!white}\textcolor{black}{0.34} &
\cellcolor{goodgreen!79!white}\textcolor{white}{0.38} & \cellcolor{goodgreen!60!white}\textcolor{white}{0.44} & \cellcolor{goodgreen!59!white}\textcolor{white}{0.39} & \cellcolor{goodgreen!20!white}\textcolor{black}{0.44} \\

\textbf{Transfusion} &
\cellcolor{goodgreen!100!white}\textcolor{white}{0.54} & \cellcolor{goodgreen!100!white}\textcolor{white}{0.56} & \cellcolor{goodgreen!64!white}\textcolor{white}{0.53} & \cellcolor{goodgreen!0!white}\textcolor{black}{0.63} &
\cellcolor{goodgreen!21!white}\textcolor{black}{0.19} & \cellcolor{goodgreen!0!white}\textcolor{black}{0.21} & \cellcolor{goodgreen!0!white}\textcolor{black}{0.18} & \cellcolor{goodgreen!0!white}\textcolor{black}{0.20} &
\cellcolor{goodgreen!31!white}\textcolor{black}{0.29} & \cellcolor{goodgreen!0!white}\textcolor{black}{0.31} & \cellcolor{goodgreen!0!white}\textcolor{black}{0.27} & \cellcolor{goodgreen!0!white}\textcolor{black}{0.38} \\

\textbf{BroadGen} &
\cellcolor{goodgreen!0!white}\textcolor{black}{0.47} & \cellcolor{goodgreen!50!white}\textcolor{black}{0.53} & \cellcolor{goodgreen!100!white}\textcolor{white}{0.54} & \cellcolor{goodgreen!93!white}\textcolor{white}{0.72} &
\cellcolor{goodgreen!0!white}\textcolor{black}{0.15} & \cellcolor{goodgreen!26!white}\textcolor{black}{0.30} & \cellcolor{goodgreen!11!white}\textcolor{black}{0.21} & \cellcolor{goodgreen!40!white}\textcolor{black}{0.37} &
\cellcolor{goodgreen!0!white}\textcolor{black}{0.22} & \cellcolor{goodgreen!34!white}\textcolor{black}{0.38} & \cellcolor{goodgreen!16!white}\textcolor{black}{0.30} & \cellcolor{goodgreen!37!white}\textcolor{black}{0.49} \\

\textbf{BroadGen(Aug)} &
\cellcolor{goodgreen!30!white}\textcolor{black}{0.49} & \cellcolor{goodgreen!0!white}\textcolor{black}{0.51} & \cellcolor{goodgreen!11!white}\textcolor{black}{0.50} & \cellcolor{goodgreen!100!white}\textcolor{white}{0.72} &
\cellcolor{goodgreen!100!white}\textcolor{white}{0.38} & \cellcolor{goodgreen!100!white}\textcolor{white}{0.54} & \cellcolor{goodgreen!100!white}\textcolor{white}{0.44} & \cellcolor{goodgreen!100!white}\textcolor{white}{0.63} &
\cellcolor{goodgreen!100!white}\textcolor{white}{0.43} & \cellcolor{goodgreen!100!white}\textcolor{white}{0.52} & \cellcolor{goodgreen!100!white}\textcolor{white}{0.47} & \cellcolor{goodgreen!100!white}\textcolor{white}{0.67} \\
\midrule
\rowcolor{gray!50}
\textbf{Oracle} &
0.32 & 0.46 & 0.44 & 0.78 &
1.00 & 1.00 & 1.00 & 1.00 &
0.48 & 0.63 & 0.61 & 0.88 \\
\bottomrule
\end{tabular}
\caption{Precision, recall and F1 scores (RRE) for all the models. Darker hues represent more favorable outcomes.}
\label{tab:all_model_wrt_oracle}
\vspace{-3mm}
\end{table*}
\addtolength{\tabcolsep}{1.5pt}

\subsection{Evaluation Framework}
\label{ss:metrics}

\subsubsection{Defining Relevance}
\label{BERT}
In eBay's Advertising ecosystem, a downstream relevance model filters the queries matched by broad match keyphrases, which is done by \textit{eBay Search}. Hence, it is crucial to avoid unnecessary recommendations that are too generic which impacts the downstream efficiency and focus on passing the subsequent filtering processes by Search on the matched queries. In addition, the queries have to be relevant to the item in terms of human judgment, or else it will not be engaged by the buyers. To assess the relevance between item and query, we employ a BERT~\cite{devlin-etal-2019-bert} cross-encoder model~\cite{wolf-etal-2020-transformers} that has been fine-tuned using judgment data derived from Large Language Models (LLMs) at eBay~\cite{dey2025judgejudgeusingllm}. This model's relevance judgment emulates human judgment while still aligning with eBay Search's definition of relevance. We use this relevance judgment to check whether a query is relevant to an item for our evaluation. Employing this cross-encoder's judgment also captures the semantic nuances of relevance that traditional lexical scoring methods generally overlook \cite{dey2025middlemanbiasadvertisingaligning}.

A wider reach for broad match keyphrases will lead to sufficient advertising impressions provided they have a competitive bid value. A good accuracy for broad match keyphrases in terms of relevant targeting will lead to better efficiency metrics like Search relevance filter pass rate for its Advertising auctions while still maintaining a healthy margin for sellers with an increased buyer engagement as a consequence. We describe achieving this desired property as maximizing the \texttt{``relevant reach''} --- \textit{A good broad match keyphrase is generally characterized by a high proportion of relevant queries in relation to the item that it reaches}. In order to quantify and measure this, we design the following two evaluation metrics.

\subsubsection{Relevant Reach Estimation (RRE)}
\label{metric:relevant_coverage}

The relevant reach of a broad match keyphrase is subject to the distribution of buyers' search queries, the broad matching mechanism, and sellers' advertising bids and budget. Although it's infeasible to compute this metric offline, we can approximate it through sampling from the Post-query distribution. In addition, we define a relevance function $\hat{R}(i, q)$ based on the definition of relevance explicated in \ref{BERT} and a broad match function B(k, q) as follows:
\begin{equation}
    \hat{R}(i, q) =
    \begin{cases}
        1 & \text{if $q$ is relevant to $i$}\\
        0 & \text{if $q$ is not relevant to $i$}
    \end{cases}
\end{equation}

\begin{equation}
    B(k, q)= 
        \begin{cases}
            1 & \text{$k$ can be broad matched to $q$}\\
            0 & \text{$k$ cannot be broad matched to $q$}
        \end{cases}\\
\end{equation}

\begin{equation}
    B@\mathcal{K}(k, q) = \mathds{1}_{\sum_{k\in \mathcal{K}}B(k,q) > 0}\\
\end{equation}
, where $k$ is a keyphrase and $q$ is a query. For broad match keywords at time $T$, the future search query distribution $Q$ is estimated by computing the search volumes of each queries from time $T$ to $T + 30$ in practice. We can now define the precision/recall type of metrics as: \vspace{-2mm}

\begin{equation}
\label{eq:RRE}
    \begin{aligned}
        Precision@\mathcal{K}(i, k) = \frac{\sum\limits_{q \sim Q}B@\mathcal{K}(k,q) \cdot \hat{R}(i,q)}{\sum\limits_{q \sim Q}B@K(k,q)}\\
        Recall@\mathcal{K}(i, k) = \frac{\sum\limits_{q \sim Q}B@\mathcal{K}(k,q) \cdot \hat{R}(i,q)}{\sum\limits_{q \sim Q}\hat{R}(i,q)}
    \end{aligned}
\vspace{1mm}
\end{equation}

 In practice, we choose the broad match criterion for $B(k,q)$ to be 1 if the tokens of $k$ are a subset of the tokens of $q$. We relax this criterion when $k$ contains N tokens ($N >= 3$) where we require $q$ to contain at least $N-1$ tokens from $k$.

\subsubsection{Proportional Token Reach (PTR)}
\label{metric:token_coverage}
While the previous metrics take care of estimating the relevant reach, a very generic broad match, by definition, can achieve the highest scores on the metrics. Hence, we also need to evaluate how specific the broad match is relative to the queries. Considering keyphrase $k$ and query $q$ as set of tokens, we define \textit{Proportional Token Reach} as: \vspace{-2mm}

\begin{equation}
\label{eq:ptr}
    PTR(k,q) = \frac{|q\bigcap k|}{|q\bigcap k|+\alpha\cdot|q-k|+\beta\cdot|k-q|}
    \vspace{1mm}
\end{equation}

The PTR computes how similar two sets of tokens are while penalizing missing tokens and additional tokens that are regulated by the scalars $\alpha$ and $\beta$, respectively. For each item, the PTR is computed for each of its recommended keyphrases and the maximum is chosen. This is averaged over all items to report the final score.

\subsection{Offline Performance}
\label{ss:performance_results}
We compare with the Transfusion and the Match Type Curation models from Section~\ref{s:related_work}. As a proxy for the match type curation strategy, we use the top pre-queries (used as input for BroadGen) as the recommended broad match type keyphrases and term it as the \textit{Prequery} model. We generate 5 broad match type keyphrases for each model. \textit{BroadGen(Aug)} indicates BroadGen with the data augmentation.

\subsubsection{RRE Results}
\label{sss:RRE_results}

We show here the RRE's precision and recall scores as defined in Section~\ref{metric:relevant_coverage} relative to the prequery model on the 4 categories of eBay in Table~\ref{tab:all_model_wrt_oracle}. We use the traditional definition of F1 on top our precision and recall scores from Equation~\ref{eq:RRE}. BroadGen(Aug) achieves the highest recall and F1 scores, despite having precision similar to the pre-query model. While generally exhibiting high precision, Transfusion has the lowest recall and F1 scores. Intuitively, recall represents the effectiveness of broad match keyphrases by measuring their ability to capture a comprehensive set of queries relevant to the seed item. Precision, on the other hand, reflects the accuracy of these broad keyphrases, indicating the overall relevance of the buyer queries matched by Search.

To better understand how does RRE fairs when there is a perfect baseline, we design an \textit{Oracle} that reports all the 1000 queries for any item as broad matched ($B(\cdot)=1$), which includes both BERT relevant and non-relevant queries leading to a 100\% recall model. Although the recall of this model is 100\%, as seen in Table~\ref{tab:all_model_wrt_oracle}, the precision is generally the lowest. This is due to the large number of queries the Oracle covers while the numerator is limited to the BERT relevant queries for each item. In essence, not every query reachable by a keyphrase is relevant to the item thus a balance between the precision and recall is essential.

\subsubsection{PTR Results}
\label{sss:ptr_results}
Table~\ref{tab:all_model_ptr_perf} shows the average PTR scores as defined in Section~\ref{metric:token_coverage} of the BroadGen(Aug) model relative to the Prequery model. A higher PTR score indicates how the keyphrases lexically resemble the queries --- it takes few modifications to a keyphrase to match a relevant query. The observations are shown with two values of $\alpha$ and $\beta$ that control the penalty for dissimilarity. For both values of $\alpha$ and $\beta$, BroadGen(Aug) performs better in all categories. When $\alpha=1.0$ and $\beta=1.5$, PTR penalizes extra words in the keyphrases more than missing words in the queries and when $\alpha=1.5,\beta=1.0$ it does vice-versa. Our model performs relatively better in the first case indicating that it requires less deletion of tokens to match the query. This is crucial as many platform's implementation of broad matching prefer not to delete tokens. BroadGen and Prequery models' similar performance in the second case suggests that our recommendations are not too generic. Transfusion being generative and not informed by any historical precedence, many of its keyphrases' tokens do not match that of the queries resulting in lower PTR.

\begin{table}[h]
\fontsize{8.2pt}{10pt}\selectfont
\centering
\begin{tabular}{c || c c || c c c c}
\toprule
\multirow{2}{*}{\textbf{Model}} & \multicolumn{2}{c||}{\textbf{Params}} & 
\multirow{2}{*}{\textbf{CAT\_1}} & 
\multirow{2}{*}{\textbf{CAT\_2}} & 
\multirow{2}{*}{\textbf{CAT\_3}} & 
\multirow{2}{*}{\textbf{CAT\_4}} \\

& $\alpha$ & $\beta$ & & & & \\
\midrule

\multirow{2}{*}{\makecell[cc]{\textbf{BroadGen}\\\textbf{(Aug)}}}
& 1.0 & 1.5 & 
\cellcolor{goodgreen!100!white}\textcolor{white}{1.16$\times$} & 
\cellcolor{goodgreen!100!white}\textcolor{white}{1.08$\times$} & 
\cellcolor{goodgreen!100!white}\textcolor{white}{1.18$\times$} & 
\cellcolor{goodgreen!100!white}\textcolor{white}{1.19$\times$} \\
& 1.5 & 1.0 & 
\cellcolor{goodgreen!85!white}\textcolor{white}{1.08$\times$} & 
\cellcolor{goodgreen!83!white}\textcolor{white}{1.01$\times$} & 
\cellcolor{goodgreen!92!white}\textcolor{white}{1.12$\times$} & 
\cellcolor{goodgreen!90!white}\textcolor{white}{1.12$\times$} \\

\midrule

\multirow{2}{*}{\textbf{Transfusion}} 
& 1.0 & 1.5 & 
\cellcolor{goodgreen!25!white}\textcolor{black}{0.64$\times$} & 
\cellcolor{goodgreen!40!white}\textcolor{black}{0.71$\times$} & 
\cellcolor{goodgreen!4!white}\textcolor{black}{0.43$\times$} & 
\cellcolor{goodgreen!14!white}\textcolor{black}{0.55$\times$} \\
& 1.5 & 1.0 & 
\cellcolor{goodgreen!23!white}\textcolor{black}{0.61$\times$} & 
\cellcolor{goodgreen!28!white}\textcolor{black}{0.67$\times$} & 
\cellcolor{goodgreen!0!white}\textcolor{black}{0.40$\times$} & 
\cellcolor{goodgreen!10!white}\textcolor{black}{0.52$\times$} \\
\bottomrule
\end{tabular}
\caption{Average PTR relative to the prequery model. Darker hues show more favorable outcomes.}
\label{tab:all_model_ptr_perf}
\vspace{-3mm}
\end{table}

\subsection{Ablation Studies}
\label{ss:ablation}

\subsubsection{Dependency on Input Queries}
\label{ss:abla_inputs}
The performance of BroadGen(Aug) in contrast to BroadGen in Table~\ref{tab:all_model_wrt_oracle} indicates the importance of data augmentation for the quality of recommendations. Across all categories, recall and F1 scores experience a sharp decline, while precision changes minimally. Although precision improves for CAT\_2 and CAT\_3, it does not significantly compensate for the drop in recall.

To enhance our understanding, we examine the percentage change in BroadGen's scores when relying solely on pre-queries. Table~\ref{tab:our_model_pq_size} illustrates a notable decline in recall and F1 scores for items with fewer than 10 input pre-queries. This situation is akin to dealing with cold items, where enough related queries are not available. With a limited number of input queries, there is insufficient data to effectively cluster and generate keyphrases. The improved performance of items with more than 10 pre-queries highlights the importance of increasing the number of queries, aided by recommendations from other recall models, which proved to be transformative for BroadGen(Aug).

\addtolength{\tabcolsep}{-1.5pt}
\begin{table}[h]
\centering
\begin{tabular}{c||ccc||ccc}
\toprule
\multirow{2}{*}{\textbf{Cats.}} & \multicolumn{3}{c||}{\textbf{\# Pre-queries $>=$ 10}}   & \multicolumn{3}{c}{\textbf{\# Pre-queries $<$ 10}}    \\ \cmidrule{2-7} 
& \textbf{Precision}       & \textbf{Recall}                        & \textbf{F1}                            & \textbf{Precision}            & \textbf{Recall}               & \textbf{F1}                   \\ \cmidrule{1-7}
CAT\_1  & +1.8\% & +72\% & +60.6\% & -0.7\% & -28.2\% & -23.7\% \\
CAT\_2 & +0.2\% & +23.3\% & +20.3\% & -0.4\% & -47.6\% & -41.3\% \\
CAT\_3 & -2.2\% & +39.1\% & +33.4\% & +2.3\% & -40.3\% & -34.3\% \\
CAT\_4  & +1.9\% & +40.1\% & +33.3\% & -1.9\% & -39.2\%  & -32.5\% \\ 
\bottomrule
\end{tabular}
\caption{Effect of number of input pre-queries on RRE metrics for BroadGen.}
\label{tab:our_model_pq_size}
\vspace{-1mm}
\end{table}
\addtolength{\tabcolsep}{1.5pt}

\definecolor{goodgreen}{RGB}{0, 90, 0}
\addtolength{\tabcolsep}{-1.5pt}
\begin{table*}[t]
\centering
\small
\begin{tabular}{
>{\columncolor[HTML]{FFFFFF}}l ||cc
>{\columncolor[HTML]{FFFFFF}}c c||cccc}
\hline
\cellcolor[HTML]{FFFFFF}                                  
& \multicolumn{4}{c||}{\cellcolor[HTML]{C0C0C0}\textbf{F1 (RRE)}} & \multicolumn{4}{c}{\cellcolor[HTML]{C0C0C0}\textbf{PTR}}    \\ 
\multirow{-2}{*}{\cellcolor[HTML]{FFFFFF}\textbf{Model - Clustering algorithm}} & \cellcolor[HTML]{FFFFFF}\textbf{CAT\_1} & \cellcolor[HTML]{FFFFFF}\textbf{CAT\_2} & \textbf{CAT\_3}                               & \cellcolor[HTML]{FFFFFF}\textbf{CAT\_4} & \cellcolor[HTML]{FFFFFF}\textbf{CAT\_1} & \cellcolor[HTML]{FFFFFF}\textbf{CAT\_2} & \cellcolor[HTML]{FFFFFF}\textbf{CAT\_3} & \cellcolor[HTML]{FFFFFF}\textbf{CAT\_4} \\
\midrule
\textbf{BroadGen(Aug) - Affinity Propagation} &
\cellcolor{goodgreen!50!white}\textcolor{black}{0.39} & \cellcolor{goodgreen!50!white}\textcolor{black}{0.49} & \cellcolor{goodgreen!0!white}\textcolor{black}{0.42} & \cellcolor{goodgreen!100!white}\textcolor{white}{0.69} &
\cellcolor{goodgreen!29!white}\textcolor{black}{0.20} & \cellcolor{goodgreen!33!white}\textcolor{black}{0.24} & \cellcolor{goodgreen!40!white}\textcolor{black}{0.22} & \cellcolor{goodgreen!27!white}\textcolor{black}{0.28} \\

\textbf{BroadGen(Aug) - Bayesian Gaussian Mixture} &
\cellcolor{goodgreen!0!white}\textcolor{black}{0.36} & \cellcolor{goodgreen!0!white}\textcolor{black}{0.45} & \cellcolor{goodgreen!20!white}\textcolor{black}{0.43} & \cellcolor{goodgreen!0!white}\textcolor{black}{0.64} &
\cellcolor{goodgreen!0!white}\textcolor{black}{0.18} & \cellcolor{goodgreen!0!white}\textcolor{black}{0.20} & \cellcolor{goodgreen!0!white}\textcolor{black}{0.18} & \cellcolor{goodgreen!0!white}\textcolor{black}{0.25} \\

\textbf{BroadGen(Aug) - Agglomerative} &
\cellcolor{goodgreen!100!white}\textcolor{white}{0.42} & \cellcolor{goodgreen!100!white}\textcolor{white}{0.53} & \cellcolor{goodgreen!100!white}\textcolor{white}{0.47} & \cellcolor{goodgreen!60!white}\textcolor{white}{0.67} &
\cellcolor{goodgreen!100!white}\textcolor{white}{0.25} & \cellcolor{goodgreen!100!white}\textcolor{white}{0.32} & \cellcolor{goodgreen!100!white}\textcolor{white}{0.28} & \cellcolor{goodgreen!100!white}\textcolor{white}{0.36} \\
\bottomrule
\end{tabular}
\caption{Comparing F1 (RRE) and PTR scores of BroadGen(Aug), when the clustering core is replaced with different algorithms. $\alpha=1.0$ and $\beta=1.5$ was used for PTR.}
\label{tab:diff_clustering}
\vspace{-2mm}
\end{table*}
\addtolength{\tabcolsep}{1.5pt}

\begin{figure*}[t]
\centering
\includegraphics[width=0.95\linewidth]{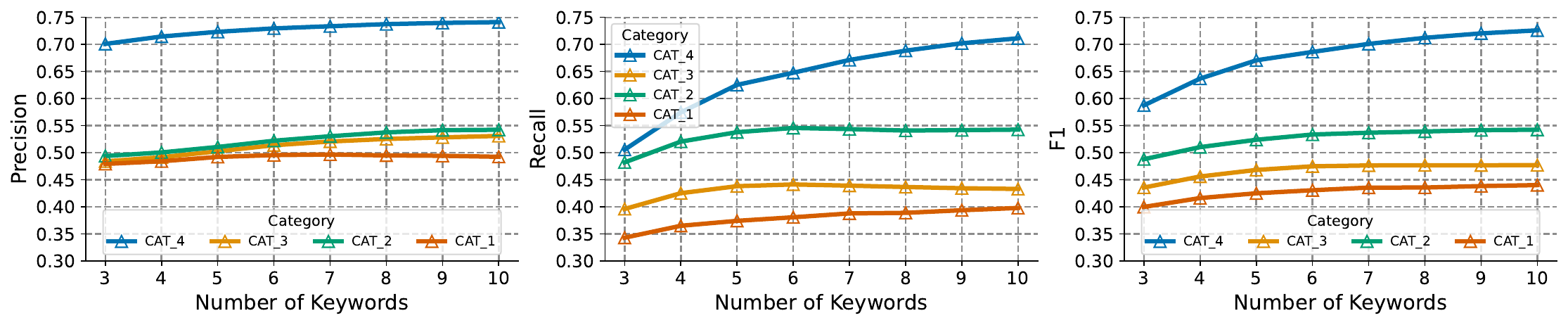} 
\caption{Comparing Precision, Recall and F1 for BroadGen(Aug) with increasing count of output keywords across the categories.}
\label{fig:precision_recall_comparison}
\vspace{-3mm}
\end{figure*}

\subsubsection{Various Clustering Algorithms}
\label{sss:clustering_algos}
Different clustering algorithms can replace the BroadGen clustering core outlined in Section~\ref{ss:clustering_core}. Algorithms like HDBSCAN~\cite{hdbscan} are unsuitable for BroadGen due to its requirement for a minimum cluster size, which can lead to grouping dissimilar queries. Table~\ref{tab:diff_clustering} compares BroadGen’s performance when its agglomerative core is replaced with Affinity Propagation~\cite{affinityPropagation} and Bayesian Gaussian Mixture~\cite{bishop-2007}. We present F1 scores based on the RRE metric alongside PTR scores. The Agglomerative method shows superior F1 scores in CAT\_1-3, albeit slightly lower in CAT\_4. Its PTR scores are notably higher than the other algorithms, as it offers greater flexibility in clustering, enabling extraction of longer keyphrases beneficial for match-type keyword recommendations. KMeans and Spectral Clustering were also evaluated, but their excessive runtime of over $140$ms and $190$ms per input on average impedes daily batch processing.

\subsubsection{Number of Keyphrases}
\label{sss:number_keyphrases}
Achieving a good balance with recall and precision for high F1 is difficult due to the limitation on the number of broad match recommendations (see Section~\ref{sss:mang_costs_revenue}) --- fewer keyphrases lead to fewer queries being reached. BroadGen(Aug) also maintains balance when increasing the number of broad match recommendations. This is evident from Figure~\ref{fig:precision_recall_comparison} where the recall as well as precision increase with the number of recommendations. The trend is notably elevated for CAT\_4, as its ambiguous characteristics lead to item diversity, resulting in a sparse distribution of relevant queries.

\subsection{Online Deployment}
\label{ss:deployment}

\subsubsection{Production Implementation}
The daily recommendations of BroadGen(Aug) is made possible through eBay's PySpark-based~\cite{apache_spark} batch inference system. The speed of our model is considerable --- with 1000 Spark executors and 4 cores, it only takes 45 minutes to finish the whole batch inference on a space of 1.5 billion items and their corresponding \textit{Queries}. 
The inputs of the production model are pre-queries with at least 5 impressions in the past 30 days augmented with exact keyphrases that pass the Bert relevance filter~\cite{dey2025judgejudgeusingllm} from two of the recall models --- fastText~\cite{fastText1, fastText2} and Graphite~\cite{ashirbad-etal-2024}. These two recall models are trained as Extreme Multi-Label Classifiers~\cite{xmc} with relevant past queries as labels for exact keyphrase recommendations. With BroadGen's processing ability, we can incorporate newer queries on a daily basis accounting for shifting label distributions (see Section~\ref{sss:shift_label_dist}) thus solving Challenge~\ref{sss:execution_requirements}.

\subsubsection{Impact}
BroadGen(Aug) was released as an additional broad match keyphrase recall for English-speaking countries over a 15-day A/B test, the treatment group received BroadGen(Aug) recommendations, while the control group did not. BroadGen(Aug) achieved a boost in broad match impressions and clicks per item by 21.73\% and 16.71\% respectively, without affecting other metrics like CTR, CVR, or ROAS showcasing it's effectiveness. Furthermore, BroadGen(Aug) increased the number of auctions that passed the search relevance filter for new broad keyphrases by 16.87\%, reflecting the improvement in efficiency of our broad match recommendations. 

\subsubsection{Sensitivity Analysis}

\begin{figure}[b]
    \centering
    \includegraphics[width=\linewidth]{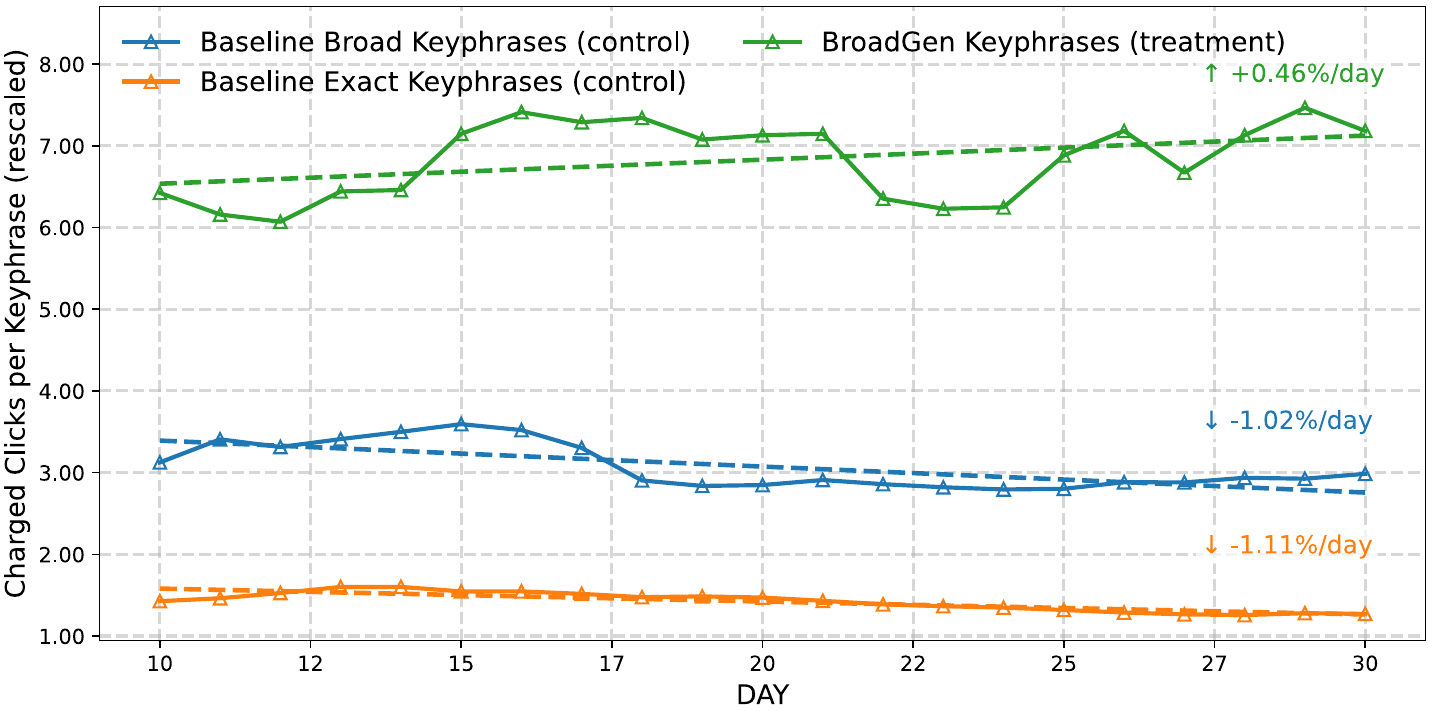}
    \caption{Performance drift over time for the BroadGen keyphrases in the treatment group vs baseline keyphrases (both exact and broad match types) from the control group.}
    \label{fig:perf_drift_plot}
\end{figure}

As noted in Section~\ref{sss:shift_label_dist}, exact-match keyphrases are vulnerable to changes in buyer search behavior, while broad-match keyphrases offer some adaptability by utilizing a flexible matching mechanism. BroadGen(Aug) advances this strategy by incorporating past queries to adjust to shifting query distributions. To evaluate its performance, we monitored BroadGen over 30 consecutive days post-advertising campaign initiation. Figure \ref{fig:perf_drift_plot} illustrates that BroadGen keyphrases consistently improve beyond the initial 10 days, encountering a minor dip after three weeks, but ultimately achieving an average daily performance increase of 0.46\%. This indicates enhanced effectiveness and stability against query drift. Conversely, in the control set, exact-match and broad-match keyphrases both exhibit a continual decline, though broad-match keyphrases fare slightly better, declining at –1.02\% daily, compared to –1.11\% for exact-match keyphrases.

\section{Conclusion and Future Work}
\label{s:conclusion}

We introduce a new framework for advancing non-exact match type recommendations, addressing the challenges associated with exact keyphrase suggestions in sponsored search advertising. Our distributed model, \textit{BroadGen}, employs a string clustering approach to offer broad match recommendations, efficiently handling up to a billion items in under an hour. Without access to ground truth match type keyphrases, we construct keyphrases by modeling the underlying token correspondence in historical queries. We also propose a unique evaluation methodology and metrics to evaluate broad match performance when real-world impressions and clicks are not available. 
In the future, we aim to expand our approach by implementing neural network equivalents to the cores of our framework. Such as, using text embeddings for input matrix generation while using a k-means or similar clustering algorithm for grouping the queries. Text summarization for the cluster representation core can be tricky as they tend to generate too generic or specific keyphrases. With the adoption of more complex cores based on LLMs we can subsequently cater to broader definitions of semantic broad match \cite{ebayMatchTypes,googleMatchTypes,semantic_approach_sigir07} and extend beyond token-based broad match in the future. Training different aspects of such a model will be difficult due to the absence of concrete supervised signals and the constraints of broad/phrase match recommendation. Moreover, the latency and infrastructure costs of these models will also be a concern due to growing number of items and queries.


\bibliographystyle{IEEEtran}
\bibliography{refs}


\end{document}